\documentclass[aps,11pt,nofootinbib, groupedaddress,preprintnumbers,superscriptaddress]{revtex4-1}

\usepackage{amssymb, amsfonts, amsmath, bm}
\usepackage[
]
{graphicx}
\usepackage{color}

\usepackage[
]
{hyperref}
\hypersetup{%
 setpagesize=false,
 bookmarksnumbered=true,%
 bookmarksopen=true,%
 colorlinks=true,%
 linkcolor=blue,%
 citecolor=red}

\begin{document}
\title{
{\Large
Chaotic particle motion around a homogeneous circular ring
}}
\author{{\large Takahisa Igata}}
\email{igata@post.kek.jp}
\affiliation{KEK Theory Center, 
Institute of Particle and Nuclear Studies, 
High Energy Accelerator Research Organization,Tsukuba 305-0801, Japan}
\date{\today}
\preprint{KEK-Cosmo-255}
\preprint{KEK-TH-2227}

\begin{abstract}
We consider test particle motion in a gravitational field generated by a homogeneous circular ring placed in $n$-dimensional Euclidean space. 
We observe that there exist no stable stationary orbits in $n=6, 7, \ldots, 10$ but exist in $n=3, 4, 5$ and clarify the regions in which they appear. 
In $n=3$, we show that the separation of variables of the Hamilton-Jacobi equation does not occur though we find no signs of chaos for stable bound orbits. 
Since the system is integrable in $n=4$, no chaos appears. 
In $n=5$, we find some chaotic stable bound orbits. 
Therefore, this system is nonintegrable at least in $n=5$ and suggests that the timelike geodesic system in the corresponding black ring spacetimes is nonintegrable. 
\end{abstract}

\maketitle

\section{Introduction}
\label{sec:1}
Circular ring structure appears in many areas of physics, 
from elementary particles to the universe. For example, planetary rings in astronomy are Newtonian gravitational phenomena in which ring shape appears clearly. 
In a strong gravity regime, the ring structure appears in an accretion disk of a compact object, and therefore, such system is modeled by a relativistic solution with a ring source~\cite{Weyl:1992, Sukova:2013jxa, Basovnik:2016awa}. In a regime where gravity is extreme, there exist ring-shaped singularities at the center of the Kerr black hole~\cite{Chrusciel:2019xuf}. Even in higher-dimensional spacetimes, which are actively studied in relativity and particle physics~\cite{Emparan:2008eg}, a ring appears as a black hole or a fundamental object such as a closed string. A typical example in relativity is a black ring spacetime, an exact solution to the 5D vacuum Einstein equation, of which the horizon topology and the central singularities are ring-shaped~\cite{Emparan:2001wn}.

In the gravitational fields of these rings, particle dynamics is basic 
for understanding the phenomena occurring in the system.
The dynamics of particles in a gravitational field generated by a homogeneous circular ring source in 3D space was numerically analyzed in detail in Ref.~\cite{Broucke:2005}; in particular, they focused on periodic orbits and classified them.
Since particle motion constrained to the 2D plane on which the ring lies is integrable because of the conservation of energy and angular momentum, the complexity of periodic orbits is relatively low. 
On the other hand, periodic orbits that deviate from the symmetric plane are relatively complicated, which leads the authors to speculate that such nature comes from nonintegrability of the system. 
Note that the integrability of this system is nontrivial and has not yet been concluded. 
Certainly, separation of variables of the equation of motion is unlikely to occur
because the Newtonian potential includes the complete elliptic integral of the first kind. 
However, since the separability of an equation of motion is a sufficient condition for its integrability~\cite{Benenti:1979}, we cannot conclude that it is nonintegrable just because it does not occur.
Answering the question of whether this system is integrable or not is one of the motivations for this study.

Now let us recall why the integrability of a test particle system is so significant.
We call a system integrable if there are as many Poisson commutable conserved quantities as or more than the system's degrees of freedom.
This nature relates to the predictability of a system because if it is nonintegrable, trajectories may exhibit chaotic behavior.
Such trajectories are generally complicated and sensitive to changes of initial conditions.
On the other hand, the predictability is preserved if a system is integrable. 
Then we can also use constants of motion to learn about the symmetry of systems and backgrounds. 
In fact, the so-called hidden symmetry of the Kerr black hole spacetime was discovered using a nontrivial constant found in the proof of the integrability of the geodesic equation~\cite{Carter:1968rr, Penrose:1973um, Floyd:1973}.
This is known today to be the fundamental quantity that characterizes the Kerr geometry.
In other words, clarifying the integrability is an effective way to discover the system's hidden symmetry.

The Newtonian potential sourced by a homogeneous circular ring in 4D space appears naturally in the Newtonian limit of the black ring solution.
The equation of motion with this potential is separable and, therefore, integrable~\cite{Igata:2014bga}. 
This property must be due to the simplicity of the potential form compared to the 3D case.
It is noteworthy that a massive particle system (i.e., timelike geodesic system) on the singly rotating black ring spacetime~\cite{Nozawa:2005eu, Hoskisson:2007zk, Igata:2010ye,Grunau:2012ai, Igata:2013be}, which restores the Newtonian potential in the weak gravity limit, exhibits chaos, i.e., the geodesic equation is nonintegrable~\cite{Igata:2010cd}.
As suggested in this example, the integrability of particle systems tends to be recovered in the Newtonian limit.
Other known examples are that the timelike geodesic system in the Schwarzschild spacetime is integrable while its Newtonian limit, the Kepler problem, is superintegrable, and that in a static dihole spacetime is chaotic~\cite{Contopoulos:1990, Contopoulos:1991} while its Newtonian limit, the Euler's 3-body problem, is integrable~\cite{Will:2008ys}.
Thus, it is quite natural to speculate that if a particle system is chaotic in the corresponding Newtonian gravitational field, the chaotic nature will also appear in a geodesic system on a relativistic gravitational field.

The Newtonian potential due to a ring source is known to have a parity of spatial dimension; 
it contains complete elliptic integrals when $n$ is odd but has a simpler structure when $n$ 
is even~\cite{Lunin:2002iz, Emparan:2009at, Igata:2020vdb}.
Based on this property and the above observations, let us make the following conjecture: 
\textit{A particle system moving in a potential sourced by a homogeneous circular ring 
in $n$-dimensional Euclidean space is nonintegrable if $n$ is odd and is integrable if $n$ is even.}
If it is true, we can predict that timelike geodesics in black ring spacetimes with an even number of spatial dimension behave chaotically.

The purpose of this paper is to verify the above conjecture. 
We use the Poincar\'e map as an indicator of chaos.
Therefore, we first identify a region where there are stable stationary orbits for each spatial dimension.
These orbits are so fundamental as to be comparable to stable circular orbits
and are important regardless of its integrability. 
Increasing an energy from the level of a stable stationary orbit, we inevitably find stable bound orbits in its vicinity.
We consider the emergence of chaotic nature by evaluating the Poincar\'e section for the stable bound orbits.

This paper is organized as follows. 
In Sec.~\ref{sec:2}, after presenting the explicit form of the Newtonian gravitational potential sourced by a homogeneous circular ring, 
we derive conditions for the existence of stationary orbits in terms of an effective potential of particle dynamics and clarify criteria for determining whether a stationary orbit is stable or unstable.
In Sec.~\ref{sec:3}, we show the regions where stable stationary orbits exist in each dimension according to the prescriptions developed in Sec.~\ref{sec:2}. 
We apply the Poincar\'e map method to stable bound orbits that appear associated with stable stationary orbits and attempt to determine the chaotic nature of particle dynamics. Section~\ref{sec:4} is devoted to a summary and discussions.

\section{Formulation}
\label{sec:2}
We consider the dynamics of a particle moving in 
a Newtonian gravitational potential sourced by a homogeneous circular ring 
in $n$-dimensional Euclidean space $\mathbb{E}^n$ ($n\geq 3$). 
Let $g_{ij}$ be the Euclidean metric, which is given by
\begin{align}
g_{ij}\:\!\mathrm{d}x^i\:\!\mathrm{d}x^j=
\mathrm{d}\zeta^2+\zeta^2 \:\!\mathrm{d}\psi^2
+\mathrm{d}\rho^2+\rho^2 
\mathrm{d}\Omega^2_{n-3},
\end{align}
where $i$, $j$ are $1, 2, \ldots, n$, and 
$(\zeta, \psi)$ are polar coordinates in 2D plane, and $(\rho, \phi_1,\ldots, \phi_{n-3})$ are spherical coordinates in the remaining $(n-2)$-dimensional space, and $\mathrm{d}\Omega_{n-3}^2$ is the metric on the unit $(n-3)$-sphere. 
Note that $0\leq \rho<\infty$ for $n\geq 4$ but $-\infty<\rho<\infty$ for $n=3$. 
Let $R$ be the radius of a homogeneous circular ring and $M$ be the total mass. 
Then the Newtonian potential sourced by the ring is given by~\cite{Igata:2020vdb}
\begin{align}
\Phi_n(\bm{r})=-\frac{GM}{(n-2) r_+^{n-2}} F\left(
\frac{1}{2}, \frac{n-2}{2}, 1; z
\right),
\end{align}
where $G$ is the gravitational constant, and 
$F$ is the hypergeometric function, and $z$ and $r_\pm$ are defined by
\begin{align}
z&=1-\frac{r_-^2}{r_+^2},
\\
r_\pm&=\sqrt{(\zeta\pm R)^2+\rho^2},
\end{align}
respectively.
The range of $z$ is restricted in $0\leq z<1$. 
Some specific forms of $\Phi_n$ for relatively small values of $n$ can be written as follows:
\begin{align}
\Phi_3(\bm{r})&=-\frac{2GM}{\pi} \frac{K(z)}{r_+},
\\
\Phi_4(\bm{r})&=-\frac{GM}{2\:\!r_+r_-},
\\
\Phi_5(\bm{r})&=-\frac{2 GM}{3\pi} \frac{E(z)}{r_+r_-^2},
\\
\Phi_6(\bm{r})&=-\frac{GM}{8(r_+r_-)^2}\left(
\frac{r_+}{r_-}+\frac{r_-}{r_+}
\right),
\\
\Phi_{7}(\bm{r})&=-\frac{2GM}{15 \pi \:\!r_-^5}\left[\:\!
-\frac{r_-^3}{r_+^3} K(z)+2\:\!\frac{r_-}{r_+} \left(1+\frac{r_-^2}{r_+^2}\right)E(z)
\:\!\right],
\\
\Phi_{8}(\bm{r})
&=-\frac{GM}{16\:\!(r_+r_-)^3} \left(
\frac{r_+^2}{r_-^2}+\frac{2}{3}+\frac{r_-^2}{r_+^2}\right),
\\
\Phi_{9}(\bm{r})
&=-\frac{2GM}{105\pi\:\!r_-^7}\left[\:\!
\left(8\:\!\frac{r_-}{r_+}+7\:\!\frac{r_-^3}{r_+^3}+8\:\!\frac{r_-^5}{r_+^5}\right)E(z)
-4\:\!\frac{r_-^3}{r_+^3}\left(1+\frac{r_-^2}{r_+^2}\right)K(z)
\:\!\right],
\\
\Phi_{10}(\bm{r})
&=-\frac{5\:\!GM}{128\:\!(r_+r_-)^4} \left(\frac{r_+^3}{r_-^3}+\frac{3}{5}\:\!\frac{r_+}{r_-}
+\frac{3}{5}\:\!\frac{r_-}{r_+}+\frac{r_-^3}{r_+^3}\right),
\end{align}
where $K(z)$ is the complete elliptic integrals of the first kind, and 
$E(z)$ is the complete elliptic integrals of the second kind.%
\footnote{
The convention of the complete elliptic integrals of the first and second kind is 
\begin{align}
K(z)=\int_0^{\pi/2} \frac{\mathrm{d}\theta}{\sqrt{1-z \sin^2\theta}},
\quad
E(z)=\int_0^{\pi/2} \sqrt{1-z\sin^2\theta} \:\!\mathrm{d}\theta. 
\end{align}}
Whether $n$ is even or odd makes considerable difference to the shape of $\Phi_n$~\cite{Lunin:2002iz, Emparan:2009at, Igata:2020vdb}.

We consider the dynamics of a particle moving in $\Phi_n$. 
Let $m$ be mass of a particle and $p_i$ be canonical momenta. The Hamiltonian is given in the form
\begin{align}
H_n
&=\frac{1}{2m}(p_\zeta^2+p_\rho^2)+V_n,
\\
\label{eq:Vn}
V_n&=\frac{L^2}{2m\zeta^2}+\frac{Q^2}{2m\rho^2}+m \Phi_n,
\end{align}
where $L=p_\psi$, and $Q^2=\gamma^{ab} p_a p_a$ (for $n\geq 4$), and $\gamma^{ab}$ is the inverse of the metric on the unit $(n-3)$-sphere, and the indices 
$a$, $b$ label $(\phi_1, \ldots, \phi_n)$. 
Both $L$ and $Q$ are constants of motion
associated with axial symmetry in the $(\zeta, \psi)$-plane and spherical symmetry in the remaining $(n-2)$-dimensional space, respectively. 
Note that in $n=3$, the term $Q^2/(2m\rho^2)$ in $V_3$ disappears. 
We call $V_n$ the effective potential in what follows. 
Since $H_n$ does not depend on time explicitly, it is also constant, 
which coincides with the conserved energy $\mathcal{E}$, i.e., 
\begin{align}
\label{eq:econs}
H_n=\mathcal{E}.
\end{align}

Now we focus on stationary orbits, i.e., particle orbits in which $\zeta$ and $\rho$ coordinates remain constant. These orbits appear when initial conditions are given to stay at an extremum point of $V_n$. 
From the equations of motion and the energy conservation law~\eqref{eq:econs}, 
we obtain the conditions for the existence of stationary orbits
\begin{align}
\label{eq:Vzeta}
\partial_\zeta V_n=0,
\\
\label{eq:Vrho}
\partial_\rho V_n=0,
\\
V_n=\mathcal{E}.
\end{align}
In the remainder of this section, 
we analyze these conditions for $n\geq 4$ and 
further provide a systematic procedure for determining the stability of stationary orbits; 
the case of $n=3$, which is formulated differently from these cases, 
will be analyzed separately in Sec.~\ref{sec:3A}. 
Solving Eq.~\eqref{eq:Vzeta} for $L^2$ and Eq.~\eqref{eq:Vrho} for $Q^2$, we obtain
\begin{align}
\label{eq:L0}
L^2&=L_0^2:=-\frac{GMm^2\zeta^3}{r_+^{n+2}}\left[\:\!
R\:\!(R^2+\rho^2-\zeta^2) \:\!
F\left(\frac{3}{2}, \frac{n}{2}, 2; z
\right)
-(\zeta+R)\:\!r_+^2 
F\left(
\frac{1}{2}, \frac{n-2}{2}, 1; z
\right)
\:\!\right],
\\
\label{eq:Q0}
Q^2&=Q_0^2:=
\frac{GMm^2 \rho^4}{r_+^{n+2}}\left[\:\!
2 R \zeta F\left(\frac{3}{2}, \frac{n}{2}, 2; z
\right)
+r_+^2 F\left(
\frac{1}{2}, \frac{n-2}{2}, 1; z
\right)
\:\!\right].
\end{align}
Note that stationary orbits exist only in the region where these 
squared angular momenta do not take negative values. 
The energy of a particle in a stationary orbit is given by
\begin{align}
\label{eq:E0}
\mathcal{E}_0=V_n|_{L=L_0, Q=Q_0}.
\end{align}

Next, let us consider conditions for stability of stationary orbits.
A stable stationary orbit is a state in which a particle on the orbit can remain in its vicinity even if a small perturbation is applied.
In this state, a particle stays at a local minimum point of $V_n$.
To determine that the extremum point of $V_n$ is a local minimum or not, 
we use the determinant and the trace of the Hessian of $V_n$,
\begin{align}
\label{eq:Hessiandet}
h(\zeta, \rho; L^2, Q^2)&=\mathrm{det} \left[\:\!
\begin{array}{cc}
\partial_\zeta^2V_n&\partial_\zeta \partial_\rho V_n
\\
\partial_\rho \partial_\zeta V_n& \partial_\rho^2 V_n
\end{array}
\:\!\right],
\\
\label{eq:Hessiantr}
k(\zeta, \rho; L^2, Q^2)&=\mathrm{tr} \left[\:\!
\begin{array}{cc}
\partial_\zeta^2V_n&\partial_\zeta \partial_\rho V_n
\\
\partial_\rho \partial_\zeta V_n& \partial_\rho^2 V_n
\end{array}
\:\!\right].
\end{align}
Using these quantities, 
we define a region $D_n$ in the $\zeta$-$\rho$ plane by 
\begin{align}
D_n=\left\{ (\zeta, \rho)\:\!|\:\! L_0^2\geq0, Q_0^2\geq0, h_0>0, k_0 >0 \right\},
\end{align}
where $h_0$ and $k_0$ are $h$ and $k$ evaluated at a stationary point, respectively, i.e., 
\begin{align}
\label{eq:h0}
h_0&=h(\zeta, \rho; L_0^2, Q_0^2),
\\
\label{eq:k0}
k_0&=k(\zeta, \rho; L_0^2, Q_0^2).
\end{align}
This gives the region where stable stationary orbits exist.

\section{Stable stationary orbits and chaotic orbits}
\label{sec:3}

\subsection{$n=3$}
\label{sec:3A}
We consider stable stationary orbits of a particle moving in the Newtonian potential $\Phi_3$. 
The explicit form of the effective potential $V_3$ is given by
\begin{align}
V_3(\bm{r})=\frac{L^2}{2m\zeta^2}-\frac{2GMm}{\pi} \frac{K(z)}{r_+},
\end{align}
where the term proportional to $Q^2$ in Eq.~\eqref{eq:Vn} does not exist. 
In this case, the condition~\eqref{eq:Vrho} takes the form 
\begin{align}
\label{eq:V3rho}
\partial_\rho V_3= \frac{2GMm}{\pi} \frac{\rho E(z)}{r_-^2 r_+}=0. 
\end{align}
Since $E(z)>0$ for $0\leq z<1$, this holds only on the symmetric plane, 
$\rho=0$. 
The condition~\eqref{eq:Vzeta} restricted on $\rho=0$ yields
\begin{align}
L^2=L^2_0
=\frac{GMm^2\zeta^2}{\pi} \left(
\frac{K(z)}{\zeta+R}+\frac{E(z)}{\zeta-R}\right),
\end{align}
where $z=4R\zeta/(\zeta+R)^2$. 
The squared angular momentum $L_0^2$ is not negative in the range
\begin{align}
\label{eq:CO3}
R<\zeta<\infty.
\end{align}
Therefore, the stationary orbits in $\Phi_3$ exist only on $\rho=0$ within the range~\eqref{eq:CO3}. Note that all of such stationary orbits are circular orbits.

Let us further restrict the inequality~\eqref{eq:CO3} to the range where stable circular orbits are allowed to exist.
We introduce the determinant $h(\zeta, \rho; L^2)$ and the trace $k(\zeta, \rho; L^2)$ of the Hessian of $V_3$ [see Eqs.~\eqref{eq:Hessiandet} and \eqref{eq:Hessiantr}] and 
define $h_0$ and $k_0$ by
\begin{align}
h_0&:=h(\zeta, 0; L_0^2)=\frac{4G^2M^2m^2 E(z)}{\pi^2 \zeta^2(\zeta^2-R^2)^2} \left(
K(z)-\frac{R^2}{(\zeta-R)^2} E(z)
\right),
\\
k_0&:=k(\zeta, 0; L_0^2)=\frac{2GMm}{\pi \zeta^2} \left(
\frac{K(z)}{\zeta+R}+\frac{E(z)}{\zeta-R}
\right),
\end{align}
respectively. Both of these are positive in the range
\begin{align}
\zeta_\mathrm{ISCO}<\zeta<\infty,
\end{align}
where $\zeta_\mathrm{ISCO}/R=1.6095\cdots$ and is determined by solving $h_0=0$~\cite{DAfonseca:2005acz,Igata:2020vdb}. 
After all, the stable stationary orbits in $n=3$ exist in the region
\begin{align}
D_3=\{\:\!(\zeta, \rho)\:\!|\:\! \rho=0,\:\! \zeta_\mathrm{ISCO}<\zeta<\infty\:\!\},
\end{align}
and all of them are stable circular orbits. We refer to $\zeta_\mathrm{ISCO}$ as 
the radius of the innermost stable circular orbit (ISCO), 
as in the case of black hole spacetimes.

\begin{figure}[t]
\centering
\includegraphics[width=15.5cm,clip]{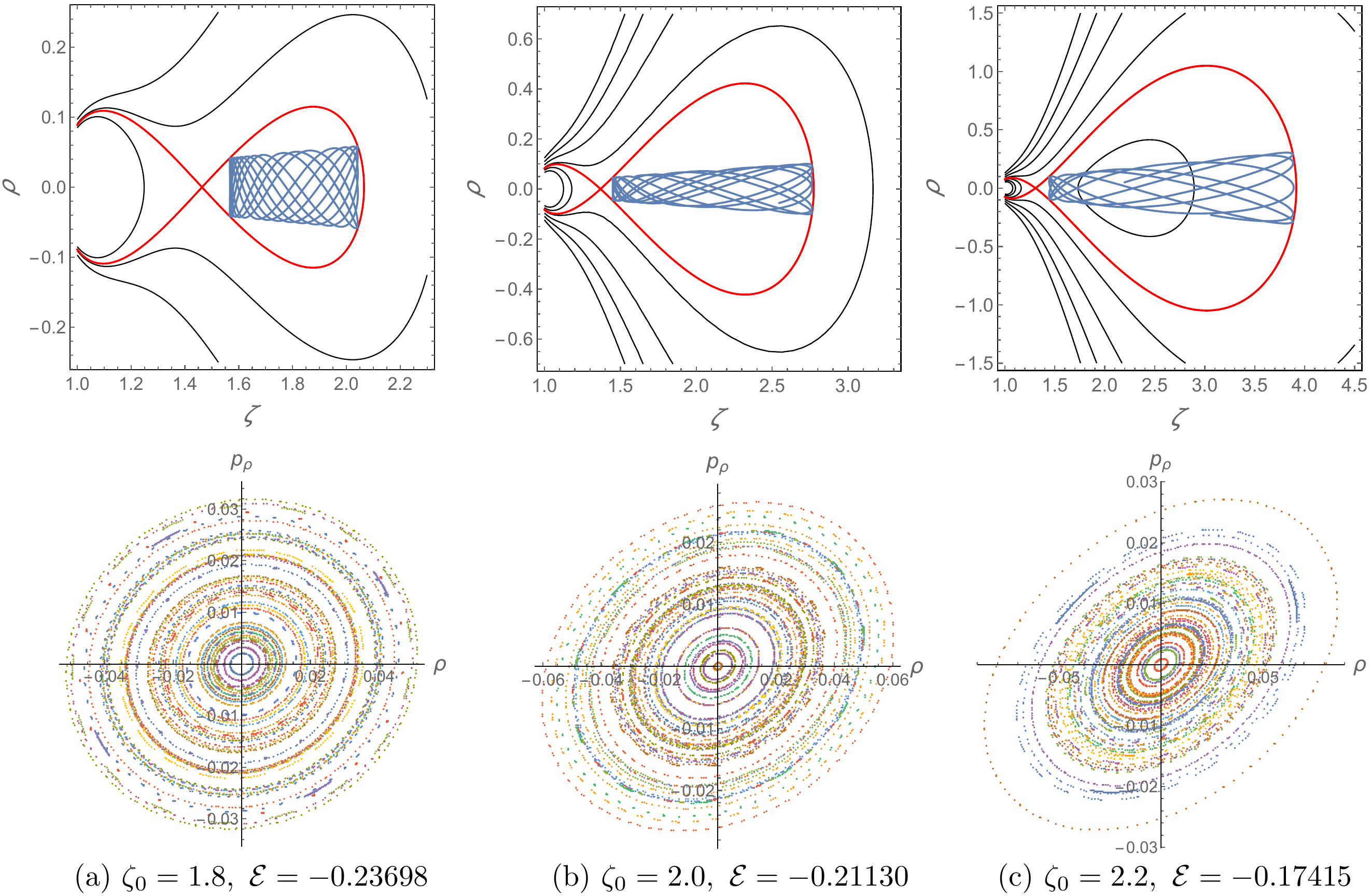}
 \caption{Typical shapes of stable bound orbits in $\Phi_3$ (upper panels) and Poincar\'e sections with the same energy and angular momenta but different initial positions and velocities (lower panels). Units in which 
$R=1$, $m=1$, and $GM=1$ are used. The local minimum point of $V_3$ is located at $(\zeta, \rho)=(\zeta_0, 0)$ in each case. 
 In the upper panels, 
the black and the red solid curves show contours of $V_3$; in particular, the red corresponds to $V_3=\mathcal{E}$. Each blue solid curve shows a stable bound orbits with energy $\mathcal{E}$.
Each point in the lower panels show a value ($\rho$, $p_\rho$) of a particle that passes through a constant-$ \zeta$ surface with $p_\zeta>0$. 
Thirty orbits with different initial conditions are superposed in each plot. 
}
 \label{fig:n=3}
\bigskip
\end{figure}

We use these results to consider the integrability of this system.
In general, when a particle in a stable stationary orbit gains some positive energy, it moves away from the local minimum of the effective potential. However, if a potential contour at an acquired energy level has still closed shape, then the particle remains confined in a finite region of the vicinity of the local minimum point. 
We call such orbits stable bound orbits. They provide information about chaotic nature of particle motion through the method of the Poincar\'e map. In Figs.~\ref{fig:n=3}, we show typical stable bound orbits in $V_3$ (upper panels) and the Poincar\'e sections (lower panels), where 
$L$ is chosen so that the local minimum point of $V_3$ coincides with the point $(\zeta, \rho)=(\zeta_0, 0)$, and $\mathcal{E}$ is chosen so that the contour of $V_3=\mathcal{E}$ (red solid curves) is closed and almost a separatrix. 
The black solid curves show the 
contours of $V_3$. 
The blue solid curves show particle trajectories with energy $\mathcal{E}$, 
which are confined inside each red closed curve. 
Though we have chosen three different parameter sets in Figs~\ref{fig:n=3}-(a)--\ref{fig:n=3}-(c), 
all of these trajectories appear to be some sort of Lissajous figures, 
which is a sign when stable bound orbits are not chaotic. 
In fact, 
the corresponding Poincar\'e sections for various initial conditions with fixing $L$ and $\mathcal{E}$ draw closed curves, as seen in the 
lower panels of Figs.~\ref{fig:n=3}, where 
the section is placed at constant-$\zeta$ plane, and phase space coordinates $(\rho, p_\rho)$ are recorded when a particle passes through the section with $p_\zeta > 0$.
Therefore, within the present analysis, we do not find any chaotic nature.

In nonintegrable systems, as is known in, e.g., 
the H\'enon-Heiles system~\cite{Henon:1964rbu}, 
the degree of chaos often increases 
if a particle in a stable bound orbit approaches a separatrix. 
It is worth noting that in our case, 
although particles approach separatrices, 
no chaos has emerged. 
However, the visualization of chaotic nature in this way may be hindered 
because the existence of the ISCO prevents stable bound orbits 
from being close enough to the ring.

The fact that signs of chaos are hard to capture may mean that this system is integrable. 
Let us discuss this possibility below. 
One of the powerful methods for analyzing the integrability of equations of particle motion is the Hamilton-Jacobi method because a sufficient condition for the integrability is the separation of variables of the Hamilton-Jacobi equation. 
It is known that the separability is closely related to the existence of the rank-2 Killing tensors~\cite{Benenti:1979}. Therefore, even in our present case, clarifying the existence of nontrivial constants of motion associated with rank-2 Killing tensors is a useful way to learn about the integrability. 
As a starting point of our discussion, we adopt a rank-2 reducible Killing tensor, i.e., a linear combination of the flat metric tensor and the symmetric tensor products of Killing vectors, 
\begin{align}
\label{eq:Kij}
K^{ij}=\alpha_0 g^{ij}+\sum_{A=1}^6\sum_{B=1}^6 \alpha_{AB}\:\! \xi_A^{(i} \:\!\xi_B^{j)},
\end{align}
where $\alpha_0$ and $\alpha_{AB}$ are constants, and 
\begin{align}
\xi_1^i&=(\partial/\partial x)^i, \quad
\xi_2^i=(\partial/\partial y)^i, \quad
\xi_3^i=(\partial/\partial z)^i, \\
\xi_4^i&=y\:\! (\partial/\partial z)^i-z\:\! (\partial/\partial y)^i, \quad
\xi_5^i=z\:\! (\partial/\partial x)^i-x\:\! (\partial/\partial z)^i, \quad
\xi_6^i=x\:\! (\partial/\partial y)^i-y\:\! (\partial/\partial x)^i
\end{align}
are the Killing vector in $\mathbb{E}_3$, which are represented by the standard 
Cartesian coordinates $(x, y, z)=(\zeta \cos \psi, \zeta \sin \psi, \rho)$. 
We assume $\alpha_{AB}=\alpha_{(AB )}$ because the antisymmetric part of $\alpha_{AB}$ does not contribute to $K^{ij}$. 
Let us focus on a quadratic quantity in $p_i$ written by $K^{ij}$ as 
\begin{align}
C=K^{ij} p_i p_j+K,
\end{align}
where $K$ is a scalar function, and without loss of generality, we have assumed that $C$ does not contain 
the first-order term of $p_i$ because, even assuming that it is included, 
it eventually disappears in the following analysis. 
In the remainder of this section, we use units in which $m=1$. 
If $C$ is a constant of motion, then 
the pair of $K^{ij}$ and $K$ must satisfy the Killing hierarchy equations~\cite{Igata:2010ny, Igata:2018dxl}
\begin{align}
\label{eq:KTeq}
&
g^{ij}\partial_i K^{kl}-K^{ij}\partial_i g^{kl}=0,
\\
\label{eq:KH2}
&\partial_i K=2 K_i{}^j \partial_j \Phi_3,
\end{align}
where $K_i{}^j=g_{ik}K^{kj}$ (see a brief review in Appendix~\ref{sec:A}). 
Our Killing tensor~\eqref{eq:Kij} is a solution to the first equation~\eqref{eq:KTeq}, which is the rank-2 Killing tensor equation in $\mathbb{E}^3$. 
Our next task is to clarify whether there is a nontrivial solution to the second equation~\eqref{eq:KH2} for $K$ with $K_{ij}$ in Eq.~\eqref{eq:Kij} as a source.
From the conditions for $K$ to be integrable, 
$\partial_{[\:\!i} \partial_{j\:\!]} K=0$, 
both $\Phi_3$ and $K^{ij}$ must satisfy the following relation: 
\begin{align}
\partial_{[\:\!i} (K_{j\:\!]}{}^k \partial_k \Phi_3)=0,
\end{align}
which leads to the restriction of the components of $\alpha_{AB}$ as
\begin{align}
\alpha_{AB}=\left[\:\!
\begin{array}{cccccc}
\alpha_{11}&0&0&\alpha_{14}&0&0\\
0&\alpha_{11}&0&0&\alpha_{14}&0\\
0&0&\alpha_{11}&0&0&\alpha_{14}\\
\alpha_{14}&0&0&0&0&0\\
0&\alpha_{14}&0&0&0&0\\
0&0&\alpha_{14}&0&0&\alpha_{66}\\
\end{array}
\:\!\right].
\end{align}
Therefore, the integrability condition for $K$ restricts 
the form of $K^{ij}$ as
\begin{align}
\label{eq:Kijred}
K^{ij}=\alpha_0 g^{ij}+\alpha_{66} \:\!\xi_6^i \xi_6^j,
\end{align}
where we have assume $\alpha_{11}=0$ because we can rescale $\alpha_0$. 
Using the restricted form~\eqref{eq:Kijred} as the source of Eq.~\eqref{eq:KH2}, we obtain
\begin{align}
K=2\alpha_0 \Phi_3,
\end{align}
where we have removed a constant term. 
Finally, we find that $C$ consists of the sum of the known conserved quantities, 
\begin{align}
C=2\alpha_0 H+\alpha_{66} L^2,
\end{align}
which is not independent from $H$ and $L^2$. 
From these results, we conclude that the separation of variables of the equation of motion does not occur. 
Note that, however, this result does not necessarily mean that 
the system is nonintegrable. 
For example, there may exist a constant of motion that is higher-order in $p_i$ more than rank-$2$~\cite{Gibbons:2011hg} or nonpolynomial form~\cite{Aoki:2016ift}. 
We need further analysis to clarify the integrability of this system, 
which is an important task for the future. 

\subsection{$n=4$}
\begin{figure}[t]
\centering
\includegraphics[width=7.0cm,clip]{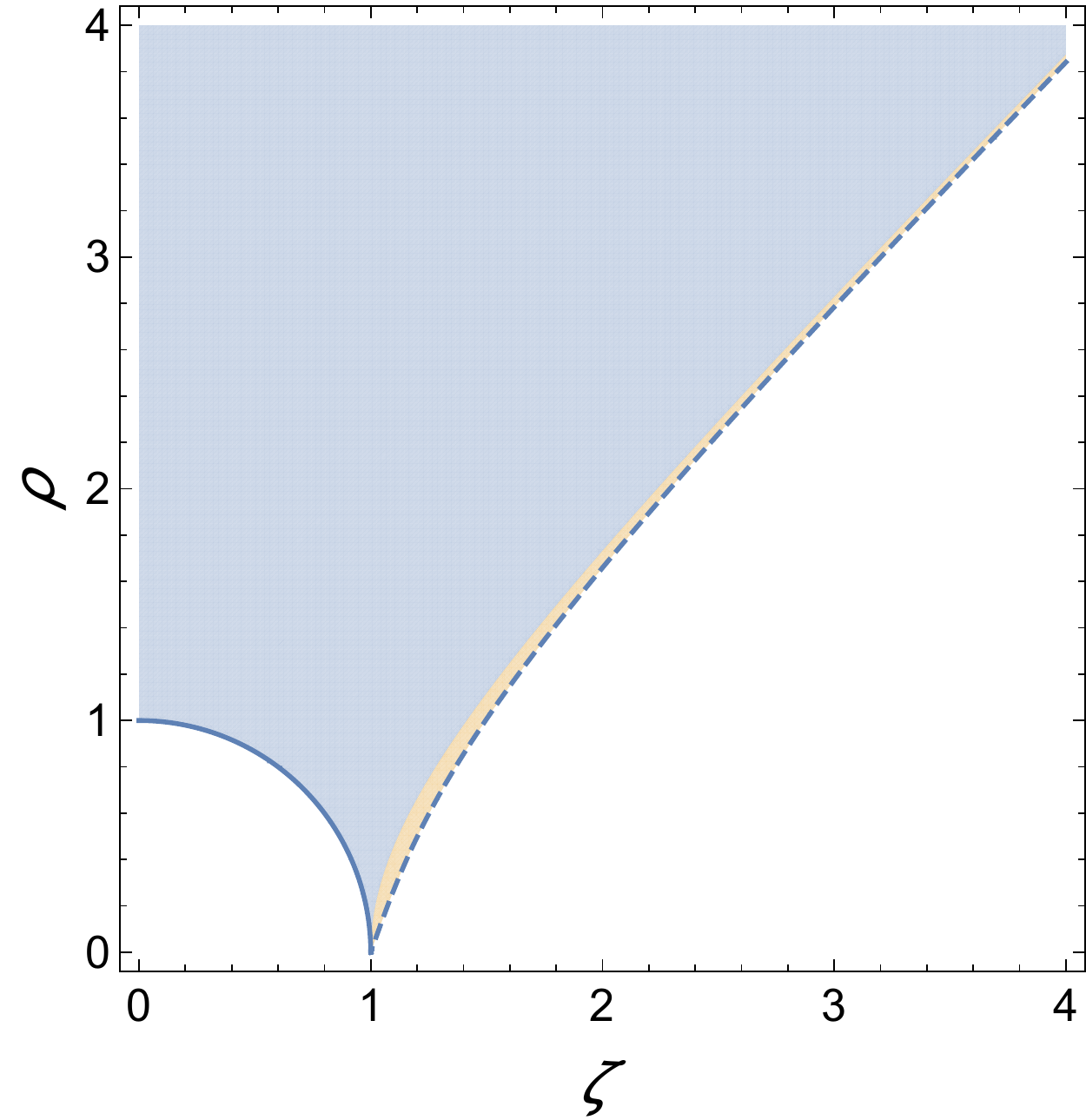}
 \caption{Region $D_4$, the allowed region for stable stationary orbits in $\Phi_4$. 
 Units in which $R=1$ are used. 
 The circular ring source is located at $(\zeta, \rho)=(1,0)$. The shaded region denotes $D_4$. 
The blue solid and dashed curve are the boundaries of $D_4$ and are determined by $L_0=0$ and $h_0=0$, respectively. 
The region $D_4$ is colored in blue when $\mathcal{E}_0<0$ and orange when $\mathcal{E}_0>0$.}
 \label{fig:n=4}
\bigskip
\end{figure}

We consider stable stationary orbits in the Newtonian potential $\Phi_4$. 
The explicit form of the effective potential $V_4$ is given by
\begin{align}
V_4=\frac{L^2}{2m\zeta^2}+\frac{Q^2}{2m\rho^2}-\frac{GMm}{2r_+r_-}.
\end{align}
As formulated in Eqs.~\eqref{eq:L0} and \eqref{eq:Q0}, 
the two squared angular momenta for stationary orbits are given by
\begin{align}
L_0^2&=GMm^2 \frac{\zeta^4(\zeta^2+\rho^2-R^2)}{r_+^3r_-^3},
\\
Q_0^2&=GMm^2 \frac{\rho^4 (\zeta^2+\rho^2+R^2)}{r_+^3r_-^3}.
\end{align}
The squared angular momentum $Q_0^2$ does not take a negative value everywhere, 
while $L_0^2$ is not negative in the range of $\zeta^2+\rho^2\geq R^2$ or $\zeta=0$, and hence only in which the stationary orbits exist. 
At the points where $L_0^2$ vanishes, 
the gravitational force in the $\zeta$ direction is just balanced.
From Eq.~\eqref{eq:E0}, the energy in stationary orbits is given by
\begin{align}
\mathcal{E}_0=-\frac{GMm R^2 (R^2+\rho^2-\zeta^2)}{2\:\!r_+^3r_-^3}.
\end{align}
Furthermore, $h_0$ and $k_0$ in Eqs.~\eqref{eq:h0} and \eqref{eq:k0} reduces to 
\begin{align}
h_0&= \frac{16 G^2 M^2 m^2 R^2}{r_+^8r_-^8}\left[\:\!
(\zeta^2+\rho^2)^2(R^2-\zeta^2+\rho^2)-R^2(R^2-\zeta^2)(R^2+\rho^2)
\:\!\right],
\\
k_0&=\frac{4 GMm (\zeta^2+\rho^2)}{r_+^3r_-^3},
\end{align}
respectively.
The positivity of $k_0$ does not make any restriction to $D_4$ because it is not negative everywhere, 
while the positivity of $h_0$ restricts $D_4$. 
Figure~\ref{fig:n=4} shows the numerical plot of $D_4$, which is drawn by the shaded region. 
The solid blue curve denotes the boundary of $D_4$ determined by $L_0=0$, i.e., 
$\zeta^2+\rho^2=R^2$,
and the dashed blue curve the boundary of $D_4$ determined by $h_0=0$. 
The region $D_4$ coincides with the region of stable stationary orbits allowed in the asymptotically far from the thin black ring in 5D spacetime; 
on the other hand, a difference appears in their vicinity~\cite{Igata:2010ye}. 
We find that a stable stationary orbit exists arbitrary close to the Newtonian ring, but in the black ring, the last stable orbit appears, which does not reach the horizon.

As was shown in Ref.~\cite{Igata:2014bga}, the Hamilton-Jacobi equation of 
this system causes the separation of variables in the spheroidal coordinate system, 
and hence this system is integrable. 
In relation to the recent work on the Newtonian analogue of the Kerr black hole~\cite{Eleni:2019wav}, our potential $\Phi_4$ is consistent with the time-time metric component 
of the 5D singly rotating Myers-Perry black hole (see Appendix~\ref{sec:B}). 
This implies that the integrability of the particle system in $\Phi_4$ 
is closely related to the integrable property 
of the timelike geodesic equation in the 5D black hole. 
Whether or not there is a further correspondence in particle dynamics, etc., other than the integrability remains an open question.

\subsection{$n=5$}
\begin{figure}[t]
\centering
\includegraphics[width=7.0cm,clip]{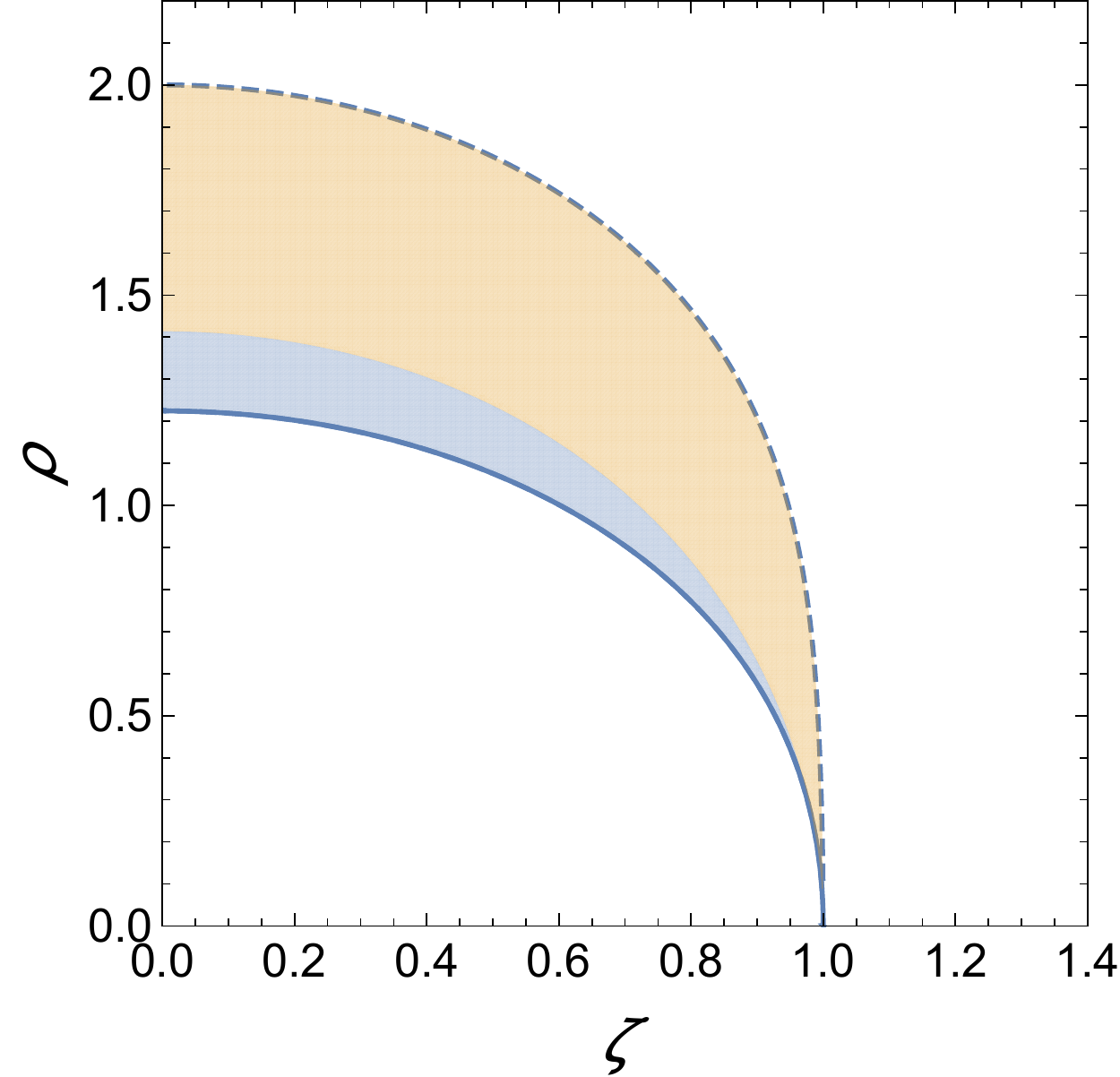}
 \caption{Region $D_5$, the allowed region for stable stationary orbits in $\Phi_5$. 
Units in which $R=1$ are used. The circular ring source is located at $(\zeta, \rho)=(1,0)$. The shaded region denotes $D_5$. 
The blue solid and dashed curves are the boundaries of $D_5$ and are determined by $L_0=0$ and $h_0=0$, respectively. The region $D_5$ is colored in blue when $\mathcal{E}_0<0$ and in orange when $\mathcal{E}_0>0$.
}
 \label{fig:n=5}
\bigskip
\end{figure}

We consider stable stationary orbits in the Newtonian potential $\Phi_5$. 
The effective potential in $n=5$ is given by 
\begin{align}
V_5=\frac{L^2}{2m\zeta^2}+\frac{Q^2}{2m\rho^2}-\frac{2GMm}{3\pi}\frac{E(z)}{r_+ r_-^2}.
\end{align}
At an extremum point of $V_5$, the squared angular momenta $L^2$ and $Q^2$ take the form 
\begin{align}
\label{eq:L0n5}
L_0^2&=\frac{GMm^2 \zeta^2}{3\pi r_+^4 r_-^3} 
\left[\:\!
r_-^2 (R^2-\zeta^2+\rho^2) K(z)-\left[\:\!
r_+^2r_-^2+8\zeta^2(R^2-\zeta^2-\rho^2)
\:\!\right] E(z)
\:\!\right],
\\
\label{eq:eq:Q0n5}
Q_0^2&=\frac{2GMm^2 \rho^4}{3\pi r_+^3r_-^4}\left[\:\!
2(r_+^2+r_-^2)E(z)-r_-^2 K(z)
\:\right],
\end{align}
respectively, and the energy $\mathcal{E}$ is 
\begin{align}
\label{eq:E0n5}
\mathcal{E}_0=-\frac{GMm}{6\pi r_-^4r_+^3}\left[\:\!
5R^4+2R^2(\rho^2-\zeta^2)-3(\zeta^2+\rho^2)^2 E(z)+r_-^2(\zeta^2+\rho^2-R^2) K(z)
\:\!\right].
\end{align}
Using Eqs.~\eqref{eq:L0n5}--\eqref{eq:E0n5} and $h_0$ and $k_0$ defined in Eqs.~\eqref{eq:h0} and \eqref{eq:k0}, 
we obtain $D_5$, where stable stationary orbits exist, as shown in Fig.~\ref{fig:n=5}.
The region $D_5$ is drawn by the shaded region, where 
the energy $\mathcal{E}_0$ is negative in blue shaded region and 
is positive in orange shaded region. 
The inner boundary of $D_5$ denoted by a solid blue curve is determined by $L_0=0$. 
Here corresponds to a balance point of the gravitational force in the $\zeta$ direction. 
The outer boundary of $D_5$ denoted by a dashed blue curve is determined by $h_0=0$.
In contrast to $D_3$ and $D_4$, which indicate unbounded regions, 
$D_5$ is distributed in a bounded region near the source.

\begin{figure}[t]
\centering
\includegraphics[width=16.0cm,clip]{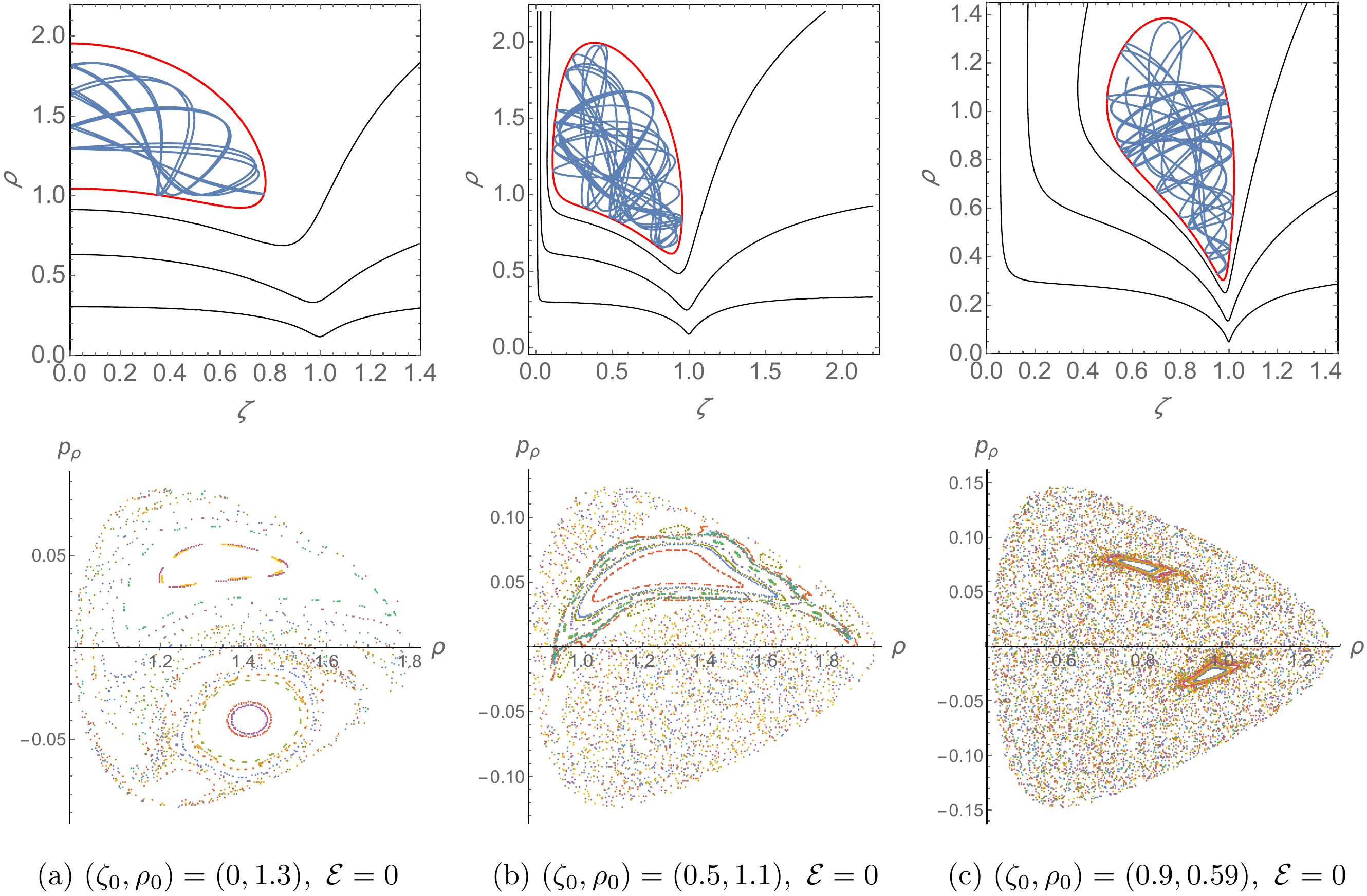}
 \caption{Typical shapes of 
 stable bound orbits in $\Phi_5$ (upper panels) and Poincar\'e sections with the same energy and angular momenta but different initial positions and velocities (lower panels). 
 Units in which $R=1$, $m=1$, and $GM=1$ are used. The values $(\zeta_0, \rho_0)$ denote the location of the local minimum point of $V_5$ in each case. 
 The black and red solid curves of each upper panel are contours of the effective potential $V_5$, which take $1, 10^{-1}, 10^{-2}$ (black), and $0$ (red). The blue curves show stable bound orbits with $\mathcal{E}=0$. The plot in the $\rho$-$p_\rho$ plane in each lower panel show the Poincar\'e sections. Thirty orbits with different initial conditions are superposed in each plot. 
 }
 \label{fig:n=5_P}
\bigskip
\end{figure}

Now, we investigate the chaotic nature of this system 
by using stable bound orbits as in Sec.~\ref{sec:3A}.
Each upper panel in Figs.~\ref{fig:n=5_P} 
draws a certain stable bound orbit (blue curve) with initial conditions at $\mathcal{E}=0$, 
where angular momenta $L$ and $Q$ are chosen so that $V_5$ 
takes a local minimum point at $(\zeta, \rho)=(\zeta_0, \rho_0)$. 
Black and red solid curves are contours of $V_5$, 
and the red is $V_5=0$. 
Each of the lower panels in Figs.~\ref{fig:n=5_P} depicts Poincar\'e sections for stable bound orbits of particles with the same $\mathcal{E}$, $Q$, and $L$ but different initial positions and velocities.
Each section is placed in a plane where $\zeta$ is constant, and phase space coordinates $(\rho, p_\rho)$ are recorded when a particle passes through the section with $p_\zeta > 0$.
In Figs.~\ref{fig:n=5_P}-(a), we find a stable bound orbit in the vicinity of the axis of symmetry, which shows a Lissajous-like pattern. However, the corresponding Poincar\'e sections show that although some of plotted points lie on closed curves on the $\rho$-$p_\rho$ plane, some of these structures are broken.
As the contour of $V_5=0$ approaches the ring such as in Figs.~\ref{fig:n=5_P}-(b), 
a stable bound orbit no longer shows a pattern like the Lissajous figure, 
and the structure of closed curves in Poincar\'e sections is broken for many initial conditions.
Their properties are more pronounced for stable bound orbits in the vicinity of the ring, such as in Figs.~\ref{fig:n=5_P}-(c).
These results indicate chaotic nature, and 
therefore, we conclude that this system is a nonintegrable system.

\subsection{$n\geq 6$}
We consider stable stationary orbits in $\Phi_n$ of the case $n\geq 6$. 
According to the prescription in Sec.~\ref{sec:2}, 
some numerical searches for $n=6, 7, \ldots, 10$ 
show that the region $D_n$ does not exist in the $\zeta$-$\rho$ plane, 
\begin{align}
\label{eq:Dn>5}
D_n=\emptyset \quad \mathrm{for} \quad n=6, 7, \ldots, 10. 
\end{align}
With this result, there are inevitably no stable bound orbits for $n=6, 7, \ldots, 10$. 
Since we cannot use the method of the Poincar\'e map without stable bound orbits, 
we need to use other criteria for determining chaos to conclude the integrability in these cases. 
The result~\eqref{eq:Dn>5} leads us to expect the absence 
of stable stationary orbits for particles in $\Phi_n$ for $n\geq 6$.

\section{Summary and discussions}
\label{sec:4}

We have considered the dynamics of particles moving in a gravitational potential sourced by a homogeneous circular ring in $n$-dimensional Euclidean space.
In each dimension below $n=11$, 
we have clarified the regions where stable stationary orbits exist.
In $n=3$, all of such orbits are stable circular orbits and exist only on the symmetric plane outside the ISCO radius, which is larger than the ring radius. 
In $n=4$, there are no stable stationary orbits on the symmetric plane, but rather in an unbounded region connected to the axis of symmetry.
In $n=5$, stable stationary orbits exist in an bounded region connected to the axis of symmetry and do not exist at infinity.
In $n=6,7, \ldots, 10$, 
no stable stationary orbits exist in whole region. 
These results would predict a region of stable stationary/bound orbits of massive particles in the far region from thin black rings in $n\geq 4$.
At least in $n=4$, the region of the existence of stable stationary orbits revealed in the Newtonian mechanics are consistent with those in the asymptotic region of the known black ring solution.

Furthermore, using stable bound orbits that appear associated with stable stationary orbits, we have analyzed chaotic nature of particle dynamics in $n=3$ and $5$, in which cases system's integrability is unknown so far. 
We have not found any chaotic nature in $n=3$ by means of the Poincar\'e map. 
It should be noted that this result does not guarantee the integrability of the system. 
However, by showing that there are no nontrivial constants of motion associated with any rank-2 Killing tensors in $\mathbb{E}^3$, at least we have clarified that the separation of variables of the Hamilton-Jacobi equation does not occur.
If this system is integrable, the proof of integrability must be achieved not by the separation of variable but by finding a constant of motion more than second-order in momentum or a non-polynomial constant.
On the other hand, in $n=5$, the Poincar\'e sections show a sign of chaos, indicating that the system is nonintegrable.

Our results suggest that the system of a freely falling particle (i.e., timelike geodesic) in 6D black ring spacetimes is nonintegrable. 
At the same time, they strongly suggest that 
there are no hidden symmetries, such as the Killing tensors. 
Therefore, finding 6D black ring solutions based on the ansatz that assumes a hidden symmetry would not work well.

Our conjecture in the introduction holds so far for $n=4$ and $5$.
We should further discuss the appearance of chaos for odd dimensions (i.e., $n=3, 7, 9,\ldots$) and should reveal integrability for even dimensions (i.e., $n=6, 8, 10, \ldots$).
Since there are various characterizations of chaos, it is important to check the chaos in several different indicators, not only in the Poincar\'e map. For example, in the current system with periodic motions, it may be useful to evaluate homoclinic trajectories and chaos analytically using the Melnikov method (see, e.g., Ref.~\cite{Polcar:2019kwu}).
This is an interesting issue for the future.

\begin{acknowledgments}
This work was supported by Grant-in-Aid for Early-Career Scientists from the Japan Society for the Promotion of Science~(JSPS KAKENHI Grant No.~JP19K14715). 
\end{acknowledgments}

\appendix

\section{Integrability condition of the Killing hierarchy equation}
\label{sec:A}
We review the condition for the existence of a constant of particle motion that is quadratic in a momentum~\cite{Igata:2010ny, Igata:2018dxl}. Let us focus on particle motion under some scalar potential force.
We use units in which particle mass $m$ is unity in this section. 
Then the Hamiltonian generally takes the form
\begin{align}
H=\frac{1}{2} g^{ij} p_i p_j+\Phi(\bm{r}),
\end{align}
where $g^{ij}$ is the inverse metric of the background space, and $p_i$ are canonical momenta, and $\Phi(\bm{r})$ is a potential. 
We introduce a dynamical quantity $C$ in the form of a second-order polynomial of momenta, 
\begin{align}
C=K^{ij} p_i p_j+K,
\end{align}
where, without loss of generality, we have assumed that $C$ does not contain a first-order term of momentum. Even assuming that it is included, that term eventually disappears in the discussion below. 

If $C$ is a constant of motion, then the Poisson bracket of $H$ and $C$ must disappear:
\begin{align}
\{H, C\}
&=\frac{\partial H}{\partial p_i} \frac{\partial C}{\partial x^i}
-\frac{\partial H}{\partial x^i} \frac{\partial C}{\partial p_i}
\\
\label{eq:PBsecond}
&=(g^{ij} \partial_i K^{kl}-K^{ij}\partial_i g^{kl})\:\! p_j p_k p_l
+(g^{ij} \partial_i K-2K^{ij} \partial_i \Phi)\:\! p_j
\\
&=0.
\end{align}
Since $p_i$ in Eq.~\eqref{eq:PBsecond} can be any value of the on-shell, the coefficients for each order of momenta must disappear. 
As a result, we obtain the Killing hierarchy equation as shown in Eqs.~\eqref{eq:KTeq} and \eqref{eq:KH2}.

\section{Newtonian analogue of a singly rotating Myers-Perry black hole}
\label{sec:B}

The metric of the Myers-Perry black hole that rotates in a single plane is given 
in the Boyer-Lindquist coordinates by
\begin{align}
\label{eq:MPsingle}
g_{\mu\nu}\:\!\mathrm{d}x^\mu\:\!\mathrm{d}x^\nu
=&-\mathrm{d}t^2+\frac{\mu}{r^{D-5}\Sigma}(\mathrm{d}t-a\sin^2\theta \:\!\mathrm{d}\phi)^2
\cr&+\frac{\Sigma}{\Delta}\:\!\mathrm{d}r^2
+\Sigma\:\!\mathrm{d}\theta^2
+(r^2+a^2) \sin^2\theta \:\!\mathrm{d}\phi^2
+r^2 \cos^2\theta \mathrm{d}\Omega^2_{D-4},
\end{align}
where $\mu$ and $a$ are mass and spin parameters, respectively, and
\begin{align}
\Sigma=r^2+a^2\cos^2\theta,
\quad
\Delta=r^2+a^2-\frac{\mu}{r^{D-5}},
\end{align}
where we use units in which $G=1$ and $c=1$ (see, e.g., Ref.~\cite{Emparan:2008eg}). 
We define a Newtonian potential $\Psi$ from the time-time component of the metric~\eqref{eq:MPsingle} as
\begin{align}
\Psi=-\frac{1+g_{tt}}{2}=-\frac{\mu}{2\:\!r^{D-5}\Sigma}.
\end{align}
In the oblate spheroidal coordinates,
\begin{align}
r&=a\:\!\xi,
\\
\theta&=\cos^{-1} \eta,
\end{align}
we obtain $\Psi$ as
\begin{align}
\Psi=-\frac{\mu}{2 \:\!a^{D-3} \xi^{D-5} (\xi^2+\eta^2)}.
\end{align}
In making the further coordinate transformation,%
\footnote{
The new coordinates are related to the Boyer-Lindquist coordinates as 
\begin{align}
\rho=\sqrt{r^2+a^2}\sin\theta,
\quad
z=r \cos\theta.
\end{align}
}
\begin{align}
z&=a\:\!\xi \eta,
\\
\rho&=a\sqrt{(1+\xi^2)(1-\eta^2)},
\end{align}
where $\xi\in [\:\!0, \infty)$ and $\eta \in [\:\!-1, 1\:\!]$, or 
equivalently, 
\begin{align}
\xi^2&=\frac{R^2+\rho^2+z^2-a^2
}{2\:\!a^2},
\\
\eta^2&=\frac{R^2-\rho^2-z^2+a^2}{2\:\!a^2},
\end{align}
where
\begin{align}
R^2=\sqrt{(\rho^2+z^2-a^2)^2+4\:\!a^2z^2}
=\sqrt{[\:\!
(\rho-a)^2+z^2
\:\!]\:\![\:\!
(\rho+a)^2+z^2
\:\!]},
\end{align}
finally we obtain the following form of $\Psi$: 
\begin{align}
\Psi=-\frac{2^{(D-7)/2}\mu}{R^2\:\! (R^2+\rho^2+z^2-a^2)^{(D-5)/2}}.
\end{align}

In $D=4$, i.e., in the case of the Kerr black hole, under the complex $\pi/2$-rotation of the parameter, $a\to i \:\!a$, 
the potential $\Psi$ corresponds to that of the Euler's 3-body problem with equal mass $m_1=m_2=M/2$~\cite{Will:2008ys}
\begin{align}
\Psi|_{D=4}=-\frac{GM/2}{\sqrt{(z+a)^2+\rho^2}}-\frac{GM/2}{\sqrt{(z-a)^2+\rho^2}},
\end{align}
where $\mu=2GM$. In the viewpoint of the separability of the equations of particle motion, the Euler's 3-body problem is closely related to the particle system of the Kerr spacetime (see recent progress in Ref.~\cite{Eleni:2019wav}). 
In $D=5$, we can find that the potential $\Psi$ reduces to the form
\begin{align}
\Psi|_{D=5}=-\frac{\mu}{2R^2}.
\end{align}
This corresponds to the Newtonian potential of a homogeneous circular ring with radius $a$ placed in the 4D Euclidean space~(see, e.g., Ref.~\cite{Igata:2020vdb}) without any complex transformation of the parameter. 
As known in Ref.~\cite{Igata:2014bga}, the equation of motion of a particle moving in this potential is integrable. This fact seems to be closely related to the integrability of the timelike geodesic equation of the 5D singly rotating Myers-Perry black hole spacetime.
For $D\geq 6$, the source that generates $\Psi$ is still an open question. 
In addition, the potential of the Myers-Perry black holes with general rotations remains unresolved.

\end{document}